\documentclass[twocolumn,aps,showpacs,prl,superscriptaddress]{revtex4-1}

\usepackage{hyperref}
\usepackage{graphicx}
\usepackage{multirow,booktabs}

\usepackage{amsmath,amssymb,latexsym,MnSymbol}
\usepackage{color}

\begin{document}
\title{Analogies between the cracking noise of ethanol-dampened charcoal and earthquakes}

\author{H. V. Ribeiro}\email{hvr@dfi.uem.br}
\affiliation{Departamento de F\'isica, Universidade Estadual de Maring\'a, Maring\'a, PR 87020-900, Brazil}
\affiliation{Departamento de F\'isica, Universidade Tecnol\'ogica Federal do Paran\'a, Apucarana, PR 86812-460, Brazil}
\author{L. S. Costa}
\author{L. G. A. Alves}
\author{P. A. Santoro}
\author{S. Picoli}
\affiliation{Departamento de F\'isica, Universidade Estadual de Maring\'a, Maring\'a, PR 87020-900, Brazil}
\author{E. K. Lenzi}
\affiliation{Departamento de F\'isica, Universidade Estadual de Maring\'a, Maring\'a, PR 87020-900, Brazil}
\author{R. S. Mendes}
\affiliation{Departamento de F\'isica, Universidade Estadual de Maring\'a, Maring\'a, PR 87020-900, Brazil}

\begin{abstract}
{ We report on an extensive characterization of the cracking noise produced by charcoal samples when dampened with ethanol. We argue that the evaporation of ethanol causes transient and irregularly distributed internal stresses that promote the fragmentation of the samples and mimic some situations found in mining processes. The results show that, in general, the most fundamental seismic laws ruling earthquakes (Gutenberg-Richter law, unified scaling law for the recurrence times, Omori's law, productivity law and B{\aa}th's law) hold under the conditions of the experiment. Some discrepancies were also identified (a smaller exponent in Gutenberg-Richter law, a stationary behavior in the aftershock rates for long times and a double power-law relationship in productivity law) and related to the different loading condition. Our results thus corroborate to elucidate the parallel between seismic laws and fracture experiments caused by a more complex loading condition that also occurs in natural and induced seismicity (such as long-term fluid injection and gas-rock outbursts in mining processes).}
\end{abstract}


\maketitle
{
Earthquakes and fractures of materials are phenomena deeply connected under the crackling noise idea~\cite{Sethna2,Salje2}, in which systems under slow perturbation respond through discrete events with a huge variety of sizes. The most fundamental seismic laws also emerge in laboratory-scale experiments related to the fracture of materials~\cite{Hirata,Diodati,Petri,Weiss,Davidsen,Kun,Kun2,Niccolini,Niccolini2,Niccolini3,Lebyodkin,Salje,Nataf,Nataf2} and have been recently reproduced by numerical discrete element simulations of porous materials~\cite{Kun_sim,Kun_sim2}. In these experiments, an external and constant loading is applied to the material and the system' response is usually obtained by recording acoustic emissions. A constant and compressive loading is considered the most suitable analogy with natural seismicity, since the main stresses underlying tectonic earthquakes are considered compressive and stationary~\cite{Main}. In fact, a very complete parallel between the acoustic emissions produced by a porous material under constant (uniaxial) compression and earthquakes was recently reported by Bar\'o \textit{et al.}~\cite{Baro}. However, there exist other important situations related to natural and induced seismicity that do not fit the previous conditions. This is the case of seismic events produced by long-term fluid injection~\cite{Langenbruch,Davidsen_intro} and gas-rock outbursts caused by the release of gas that is common and represent a serious threat in coal mining~\cite{Kong,Meng}. 

In these situations (where the loading is internal, transient and irregular), a complete parallel between the cracking noise of materials and the fundamental seismic laws has not been established yet, despite the considerable interest in mining processes. Here we design a simple experiment that captures the previous features. Specifically, we study the acoustic emissions of charcoal samples dampened with ethanol. At room temperature, we observe that the ethanol is absorbed through the pores of the samples and soon evaporates, creating different and irregularly distributed internal stresses that promote the fragmentation of samples and somehow mimic the situations found in mining. We show that these acoustic events fulfill the Gutenberg-Richter law~\cite{Knopoff_intro,Utsu_intro,Kagan_intro,Godano_intro} (with a power-law exponent smaller than those reported for earthquakes) and the unified scaling law for the recurrence times between events~\cite{Bak_intro,Corral_intro0,Corral_intro,Corral_intro2,Saichev_intro,Touati_intro} (with parameters very close to those reported for small mine-induced seismicity~\cite{Davidsen_intro}). We also characterize the sequence of aftershocks/foreshocks, where the Omori's decay~\cite{Utsu_intro2,Shcherbakov_intro,Sornette_intro} is observed to hold only for short times ($\sim$6 seconds), from which these rates display a stationary behavior. Still on the aftershock sequences, we investigate the productivity law~\cite{Helmstetter_intro}, where a double power-law relationship between the number of aftershocks and the energy of the triggering mainshock is found (the first power-law exponent is much smaller than those reported for earthquakes, while the second is in the range of earthquakes). We also find that the relative difference in energy magnitude between the mainshock and its largest aftershock approaches the value of $1.2$ as the mainshock energy increases (that is, an approximate quantitative agreement with the B{\aa}th's law~\cite{Helmstetter_intro2}). Thus, in general, we verify that the fundamental seismic laws hold in a fracture experiment caused by an internal, non-stationary and irregular loading; however, our results also reveal some significant differences (a smaller exponent in Gutenberg-Richter law, a stationary behavior in the aftershock rates for long times and a double power-law relationship in productivity law), which we attribute to the different loading condition.
}

\begin{figure}[!hb]
\includegraphics[scale=0.22]{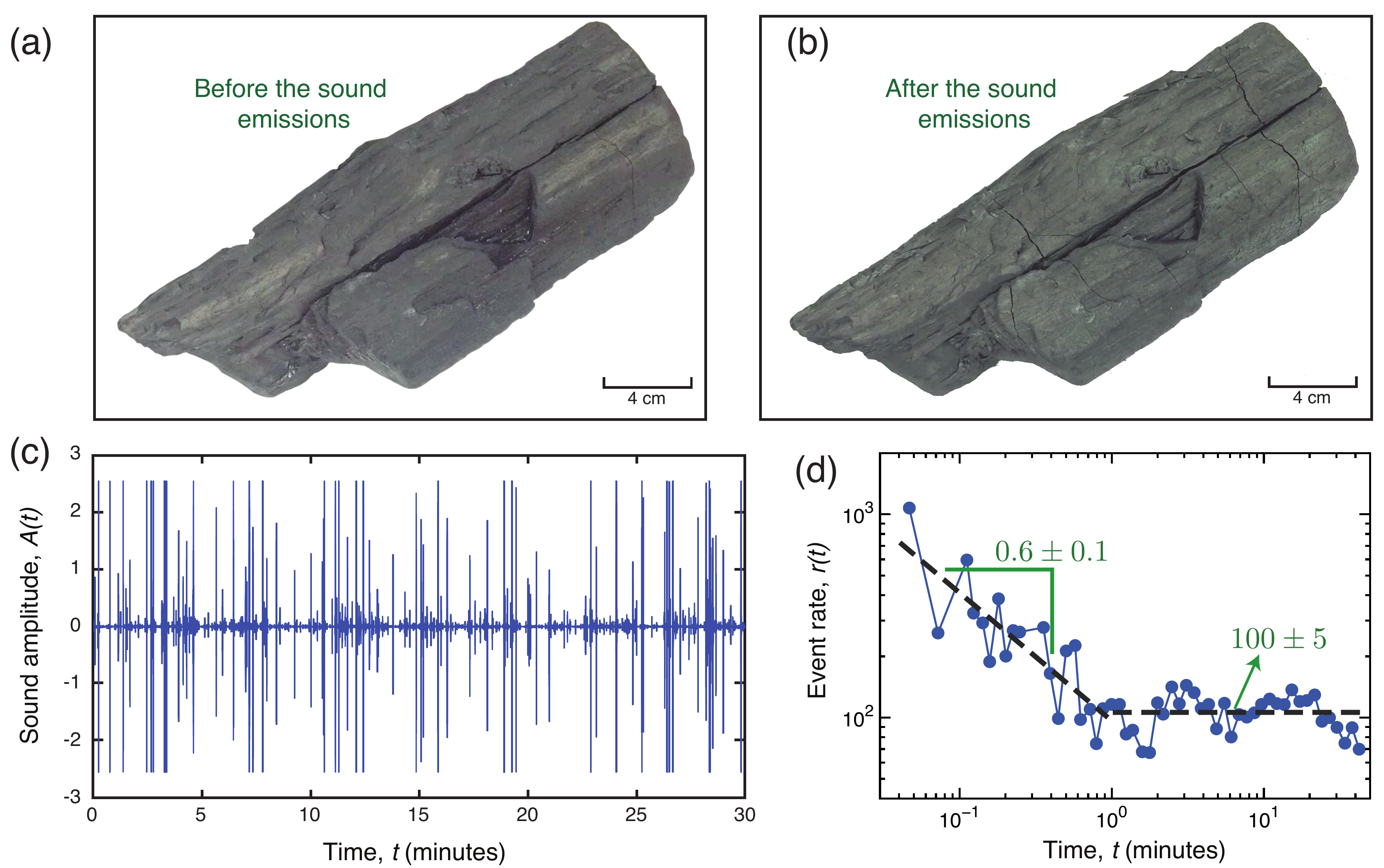}
\caption{Schematic description of the experiment. { Figures (a) and (b) show pictures of a sample (Sample \#2) immediately before and at the end of the acoustic emissions (see a video of this process and a selection of large events at \url{http://youtu.be/8Y_uJWVeU7w} and \url{http://youtu.be/ILCMgFtQjnc}). In (b), we note  several fissures caused by the cracking process. Figure (c) shows the normalized sound amplitudes $A(t)$ recorded during this process. Figure (d) shows that the rate of activity $r(t)$ (number of events per minute) displays an initial power-law decay ($r(t)\sim t^{-0.6\pm0.1}$, for $t<1$~min) followed by a nearly stationary behavior with average $\langle r\rangle = 100\pm5$ (see also Fig.~S1).}
}
\label{fig:1}
\end{figure}

{ In the experiment, charcoal samples (Fig.~\ref{fig:1}a) for domestic use ($\sim\!\!200$~g), made of \textit{Eucalyptus sp.}, are dampened with $\sim\!\!30$~ml of ethanol (for domestic use, hydrated with 7\% water), and most of it is absorbed through pores of the samples. The results we report are based on six samples of about the same size. After $\sim\!\!5$ minutes, the samples start to produce the cracking noise that is recorded by a condenser microphone (Shure Microflex MX202W/N) positioned at $\sim\!\!20$~cm from the samples with a sampling rate of $48$~kHz. The samples emit sound for $\sim\!\!30$ minutes and during the processes they also crack, ending-up very fragmented (Fig.~\ref{fig:1}b). Figure~\ref{fig:1}c shows the normalized sound amplitudes (that is, the original amplitudes divided by the aggregated standard deviation), $A(t)$, recorded during this process. Notice that the sound emissions occur in discrete events with different magnitudes. The normalized energy associated to an event $i$ is evaluated (analogously to other fracture experiments) via $E_i= \int_{t_{\text{ini}_i}}^{t_{\text{end}_i}} I(t) dt$, where $I(t) = \frac{A^2(t)}{\text{max}[A^2(t)]}$ and $t_{\text{ini}_i}$ ($t_{\text{end}_i}$) represents the start (end) time of the event. The value of $t_{\text{ini}_i}$ is chosen as the time for which $I(t)$ initially exceeds a threshold $I_{\text{min}}$, whereas $t_{\text{end}_i}$ represents the time for which $I(t)$ stays below $I_{\text{min}}$ for more than $\Delta t$ seconds. All results presented here were obtained with $I_{\text{min}}=10^{-5}$ and $\Delta t=0.1$; however, different values for these parameters (we have tested for $I_{\text{min}}$ from $10^{-5}$ to $6\times10^{-4}$ and $\Delta t$ from $0.025$ to $0.5$) do not change our results. The location time associated with an event $i$ is defined as $t_i = (t_{\text{end}_i} - t_{\text{ini}_i})/2$. We have verified that the rate of activity $r(t)$ (number of events per minute) displays an approximate power-law decay in the beginning of the process followed by a nearly stationary behavior (Fig.~\ref{fig:1}d). In our analysis, we have dropped out the 5\% initial and final events in order to keep the activity rates nearly stationary (Fig.~S1).
}


\begin{figure}[!ht]
\includegraphics[scale=0.27]{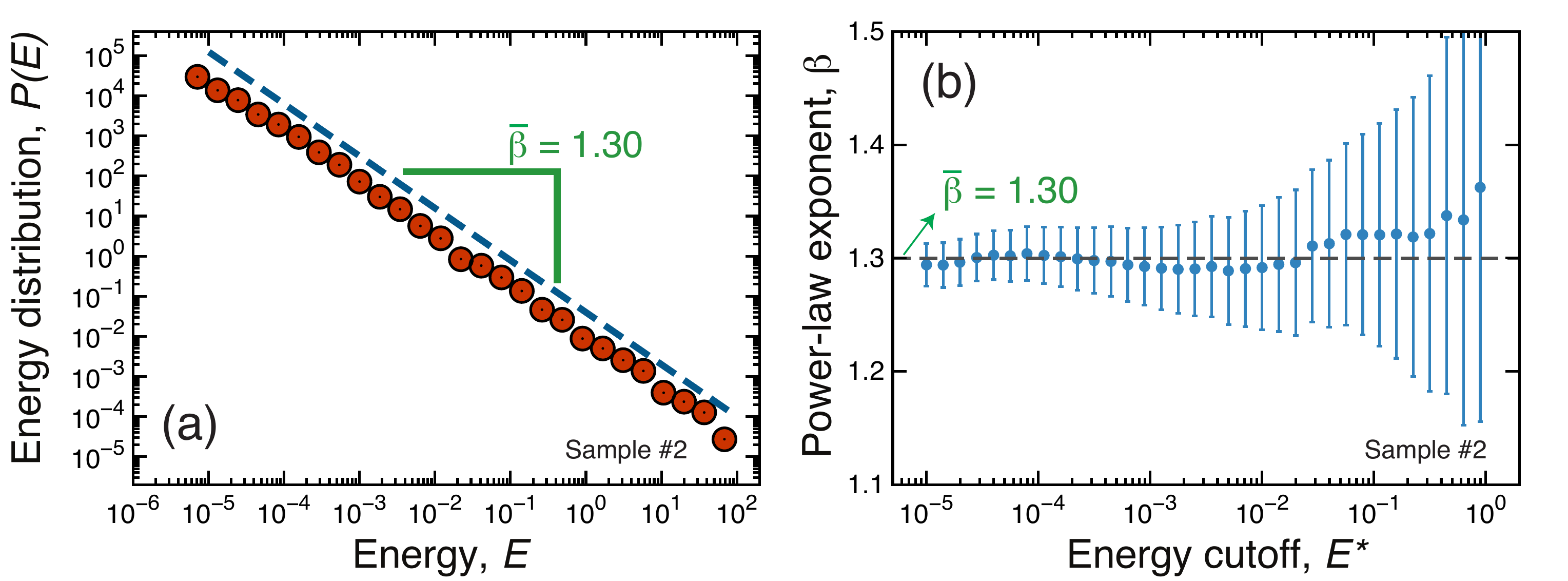}
\caption{The Gutenberg-Richter law. Figure (a) shows the probability distribution of the energies $E$ using data from one experiment (Sample \#2). The dashed line represents a power-law decay where $P(E)\sim 1/E^{\bar{\beta}}$ with $\bar{\beta}=1.30$. Figure (b) shows the values of the power-law exponents $\beta$ obtained via maximum likelihood method as a function of a lower energy cutoff $E^*$, that is, considering only events with $E> E^*$. The error bars are $95\%$ bootstrap confidence intervals. The value of $\bar{\beta}$ is the weighted average of $\beta$ over $E^*$, where the weights are chosen to be inversely proportional to the lengths of the confidence intervals. The results for other samples are very similar (Figs.~S2 and~S3).
}
\label{fig:2}
\end{figure}

We start by evaluating the probability distribution for the energies $E$. Figure~\ref{fig:2}a shows this distribution for one of the samples, where it exhibits a remarkable power-law behavior compatible with the Gutenberg-Richter law, that is, $P(E)\sim1/E^\beta$, over several decades. Figure~\ref{fig:2}b shows the values of $\beta$ estimated via maximum likelihood method for different low energy cutoffs $E^*$. We note that $\beta$ is quite stable over $E^*$. In order to assign a characteristic exponent ($\bar{\beta}$) to each sample, we have evaluated the average of $\beta$ over $E^*$. The six samples yield values for $\bar{\beta}$ in the range [1.27--1.33], which are smaller than the $\beta\approx1.67$ observed for earthquakes~\cite{Knopoff_intro,Utsu_intro,Kagan_intro,Godano_intro} and in the Bar\'o \textit{et al.}~\cite{Baro} experiment ($\beta\approx1.40$). 


\begin{figure}[!ht]
\includegraphics[scale=0.27]{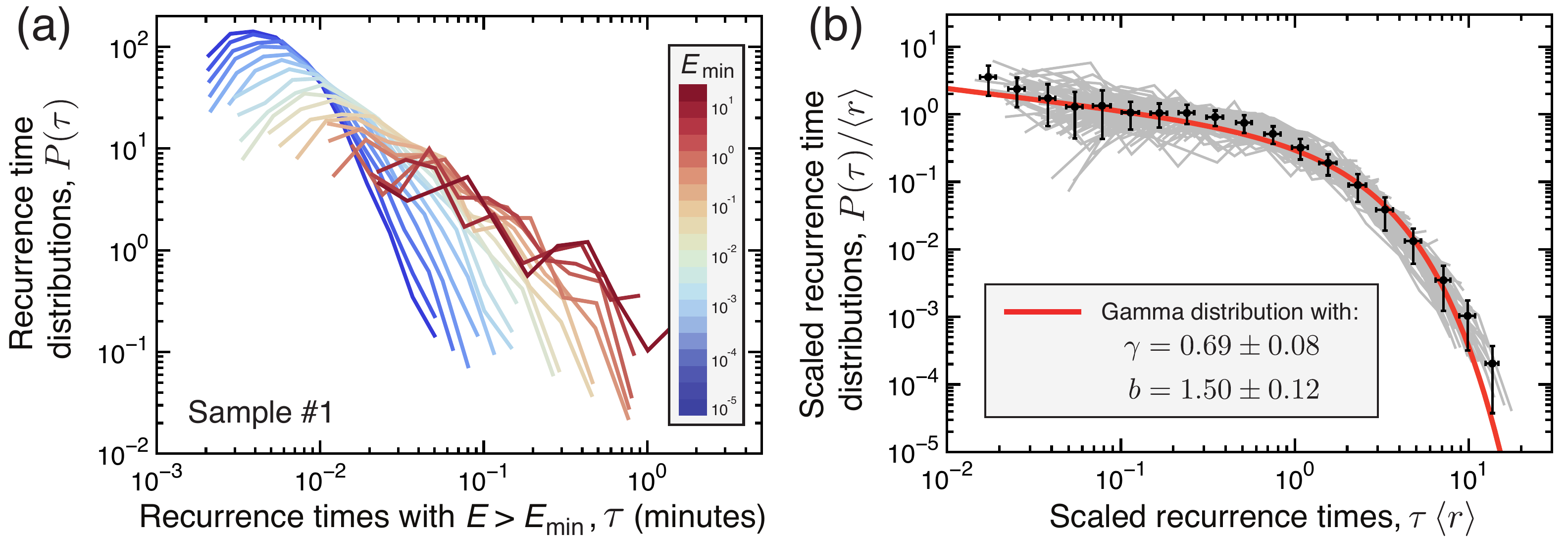}
\caption{Self-similarity of the recurrence times and the universal scaling law. Figure (a) shows the probability distributions of recurrence times $\tau$ with $E>E_{\text{min}}$ using data from one experiment (Sample \#1). Each curve is associated with a value of $E_{\text{min}}$, as indicated by the color code. Figure (b) shows the distributions rescaled by the mean rates of activity $\langle r\rangle$ (gray lines) using data from all experiments (Samples \#1-6). The black circles are the window average over all distributions and the error bars are $95\%$ bootstrap confidence intervals. The solid red line is the gamma distribution adjusted to the average distribution via ordinary least squares method (the parameters are shown in the plot). Similar results are observed for each sample (Figs.~S4 and~S5).}
\label{fig:3}
\end{figure}

Another important aspect of earthquakes is related to the time intervals $\tau$ between events above a lower bound energy $E_\text{min}$ (also called recurrence or waiting times). Bak \textit{et al.}~\cite{Bak_intro} have proposed that after accounting for the spatial location of the events, the distributions of $\tau$ collapse into a single curve. Corral~\cite{Corral_intro0,Corral_intro,Corral_intro2} has argued that the occurrence of earthquakes differs from region to region and proposed an extension to the Bak \textit{et al.} procedure, by including the local rates of seismic activities $r_{xy}$ in the scaling operation. Thus, the distributions of $\tau$ become self-similar, $P(\tau) = r_{xy} f(r_{xy} \tau)$, where $f(x)$ is a scaling function. When $r_{xy}$ is time-dependent, $f(x)$ exhibits different power-law regimes that are almost universal across several different seismic regions; whereas for $r_{xy}$ nearly stationary, $f(x)$ is usually adjusted by a gamma distribution, $f(x)\propto x^{\gamma-1} \exp(-x/b)$, where $\gamma$ and $b$ are fitting parameters. { In fact, as proposed by Saichev and Sornette~\cite{Saichev_intro}, this behavior is an emergent property of aftershock superposition that holds in real seismicity under certain conditions~\cite{Touati_intro}}.
Figure~\ref{fig:3}a shows the distributions of $\tau$ obtained from one of the samples and by considering several values of $E_\text{min}$, where it is clear that $P(\tau)$ depends on $E_\text{min}$. Figure~\ref{fig:3}b shows the same distributions (for all samples) rescaled by the mean rates of activity $\langle r \rangle$. { We observe a good collapse of the distributions and that the gamma distribution is a reasonable fit to the average behavior with $\gamma=0.69\pm0.08$ and $b=1.50\pm0.12$. These values are  very close to those reported for earthquakes ($\gamma=0.67\pm0.05$ and $b=1.58\pm0.15$~\cite{Corral_intro}) and small mine-induced seismicity ($\gamma=0.74\pm0.02$ and $b=1.35\pm0.06$~\cite{Davidsen_intro}) --- see also Refs.~\cite{Davidsen,Niccolini,Niccolini2,Niccolini3}}. 

\begin{figure}[!ht]
\includegraphics[scale=0.27]{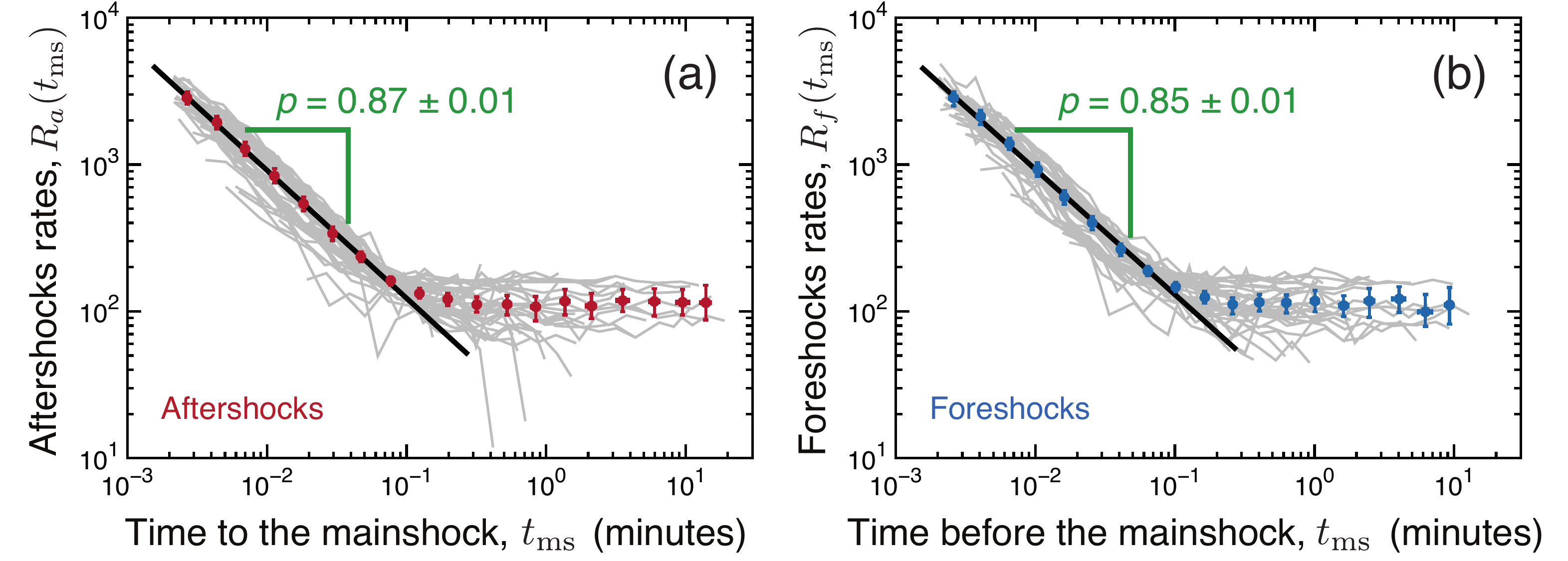}
\caption{The Omori's law for aftershocks and foreshocks. Figure (a) shows the number of aftershocks per unit of time (aftershock rates), $R_a(t_{\text{ms}})$, as function of the time to mainshock, $t_{\text{ms}}$, employing data from all experiments (Samples \#1-6). Figure (b) shows the analogue plot for the foreshocks (foreshock rates, $R_f(t_{\text{ms}})$, versus the time before the mainshock, $t_{\text{ms}}$). The mainshocks have been defined as events with energy in the range $10^{j}$--$10^{j+1}$, with $j=-4,-3,\dots,3$. Each gray curve is an Omori's plot for one of the samples with the mainshocks in one of the energy ranges. Figures~S6 and S7 show the results for each sample, identifying the energy range of the mainshocks. In both plots, the circles (red for aftershocks and blue for foreshocks) are the window average over all curves and the error bars are $95\%$ bootstrap confidence intervals. We observe power-law decays (of about two decades) for $R_a(t_{\text{ms}})$ and $R_f(t_{\text{ms}})$ for $t_{\text{ms}}<0.1$~min followed by a plateau-like behavior. The black lines are power-law functions, $R_{\{a,f\}}(t_{\text{ms}})\sim t_{\text{ms}}^{-p}$, adjusted to the average behaviors for $t_{\text{ms}}<0.1$~min via ordinary least squares method. The power-law exponents $p$ are shown in the plots.
}
\label{fig:4}
\end{figure}

We now focus on quantifying the Omori's law in our data. {The Omori's law~\cite{Utsu_intro2,Shcherbakov_intro,Sornette_intro} establishes that the number of aftershocks per unit of time $R_a(t_{\text{ms}})$ decays as a power-law function of the elapsed time since the mainshock $t_{\text{ms}}$, that is, $R_a(t_{\text{ms}})\sim 1/t_{\text{ms}}^p$. The value of $p$ for earthquakes differs from one catalog to another (probably due to different tectonic conditions), usually lying in the range [0.9--1.5]~\cite{Utsu_intro2}; its value also depends on the magnitude of the mainshock~\cite{Sornette_intro}.  In fracture experiments, Hirata~\cite{Hirata} showed (for basalt) that the value of $p$ decreases during the fracturing process and Bar\'o \textit{et al.}~\cite{Baro} reported a quite stable Omori's decay (about six decades) with $p=0.75\pm0.10$. In our case, we define the mainshock events as those with energy $E_{\text{ms}}$ in the range $[10^{j}$--$10^{j+1}]$ (with $j=-4,-3,\dots,2$) and a sequence of aftershocks is the events following the mainshock until another mainshock event is found. We calculate the aftershock rates $R_a(t_{\text{ms}})$ as a function of the elapsed time since the mainshock, $t_{\text{ms}}$, averaging over all events in the same energy window. Figure~\ref{fig:4}a shows $R_a(t_{\text{ms}})$ for all energy ranges and samples that we analyzed. Our results show that the Omori's decay (of about two decades) with $p=0.87\pm0.01$ only holds for short times ($t_{\text{ms}} \lesssim0.1$~min), from which a stationary behavior is observed for $R_a(t_{\text{ms}})$. This stationary behavior indicates that late aftershocks occur randomly in time, such as in stochastic process with no memory. Very similar results are obtained for the foreshock rates (Fig.~\ref{fig:4}b). 
}

\begin{figure}[!ht]
\includegraphics[scale=0.27]{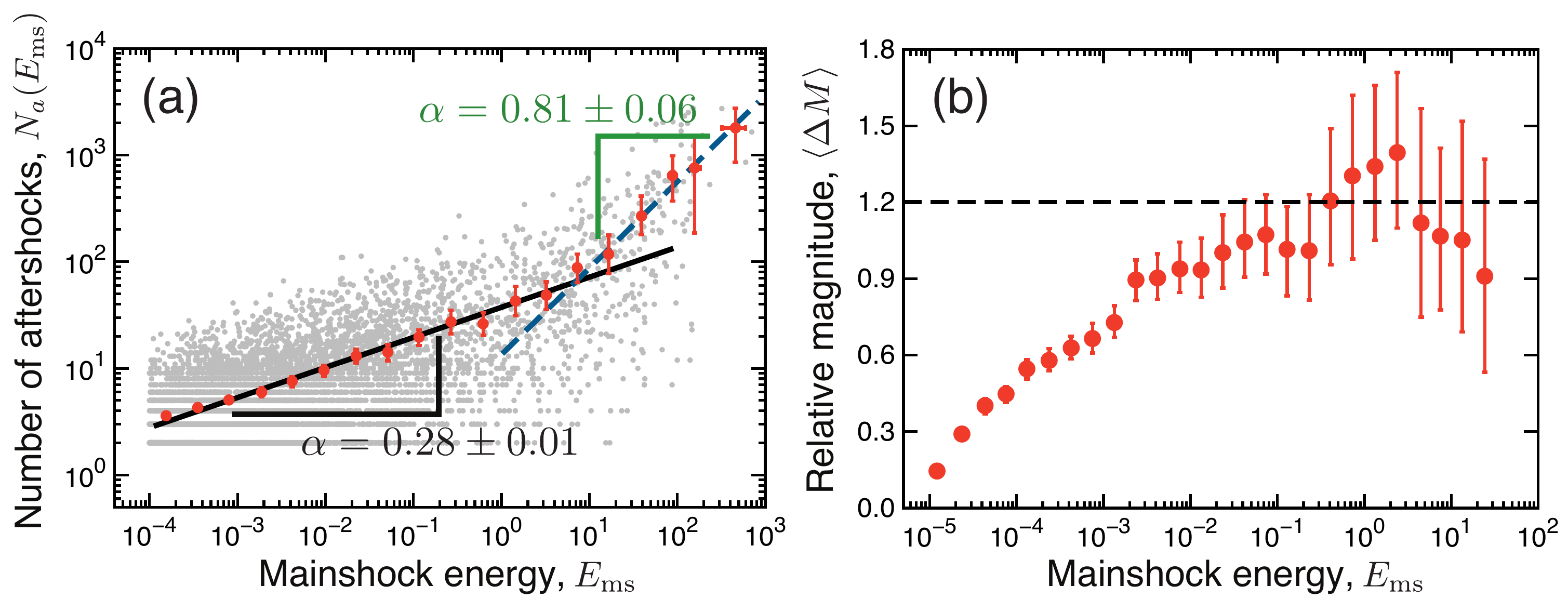}
\caption{The productivity law and the B{\aa}th's law. The gray dots in (a) show the number of aftershocks $N_{a}(E_{\text{ms}})$ triggered by a mainshock of energy $E_{\text{ms}}$ (in log-log scale). We have aggregated data from all experiments (Samples \#1-6) and the results by samples are quite similar (Fig.~S8). The red circles are window averages and the errors bars are $95\%$ bootstrap confidence intervals. For $E_{\text{ms}} <10$, the average behavior of $N_{a}(E_{\text{ms}})$ is adjusted (via ordinary least squares) by a power-law relationship $N_{a}(E_{\text{ms}})\sim E_{\text{ms}}^\alpha$, where $\alpha=0.28\pm0.01$ (black solid line). For greater mainshocks ($E_{\text{ms}} >10$), the average behavior starts to deviate from the previous power-law and can be adjusted by another power-law with $\alpha=0.81\pm0.06$ (blue dashed line). Figure (b) shows the average value of the relative difference in magnitude ($\langle\Delta M \rangle$) between the mainshock, $\log E_{\text{ms}}$, and its largest aftershock, $\log E_{\text{la}}$, as a function of the mainshock energy $E_{\text{ms}}$ (the $x$-axis is in log scale). The red circles are the results obtained with data from all experiments (individual samples show a very similar behavior, see Fig.~S9) and the error bars are $95\%$ bootstrap confidence intervals. Notice that $\langle\Delta M \rangle$ approaches the value of $1.2$ (dashed line) as $E_{\text{ms}}$ increases.
}
\label{fig:5}
\end{figure}

{ The productivity law states that the number of aftershocks $N_a(E_{\text{ms}})$ triggered by a mainshock of energy $E_{\text{ms}}$ is related to $E_{\text{ms}}$ via  $N_a(E_{\text{ms}})\sim E_{\text{ms}}^\alpha$, with $\alpha\approx0.8$ for earthquakes~\cite{Helmstetter_intro}. In order to quantify this law in our experiment, we count the number of aftershocks $N_{a}(E_{\text{ms}})$ that a mainshock of energy $E_{\text{ms}}$ triggers.} Figure~\ref{fig:5}a shows $N_{a}(E_{\text{ms}})$ versus $E_{\text{ms}}$ for all aftershock sequences (defined as in the Omori's analysis) in log-log scale, where (despite the scatter) a significant dependence is observed (Pearson correlation of $\approx0.7$). This figure also shows the window average of these data, from which the relationship between $N_{a}(E_{\text{ms}})$ and $E_{\text{ms}}$ becomes clear: for $E_{\text{ms}}<10$, we have $N_{a}(E_{\text{ms}})\sim E_{\text{ms}}^\alpha$, with $\alpha=0.28\pm0.01$; whereas for $E_{\text{ms}}>10$, we find another power-law with $\alpha=0.81\pm0.06$. Thus, the first power-law exponent is similar to that reported by Bar\'o \textit{et al.}~\cite{Baro} ($\alpha\approx0.33$), while it is smaller than the ones reported for earthquakes ($\alpha \in$ [$0.7$--$0.9$]~\cite{Helmstetter_intro}) and also for creep of ice single crystals ($\alpha\approx0.6$~\cite{Weiss}). On the other hand, the second power-law exponent is analogous to earthquakes, 
but it should be considered much more carefully because it only accounts for about $3\%$ of the data in a region where the relationship $N _{a}(E_{\text{ms}})$ versus $E_{\text{ms}}$ displays a large scatter. 

{ Still on the aftershock sequences, we address the B{\aa}th's law~\cite{Helmstetter_intro2}, which states that the relative difference in energy magnitude (that is, $\log E$) between the mainshock and its largest aftershock is (on average) close to $1.2$, regardless of the mainshock magnitude.} To do so, we calculate the relative difference in energy magnitude between a mainshock and its largest aftershock ($\Delta M = \log E_{\text{ms}} - \log E_{\text{la}}$, where $E_{\text{la}}$ is the energy of the largest aftershock) as a function of mainshock energy $E_{\text{ms}}$. Figure~\ref{fig:5}b shows the average (over all samples) of this relative magnitude $\langle \Delta M \rangle$ as function of the $E_{\text{ms}}$. We note that $\langle \Delta M \rangle$ is systematically smaller than $1.2$ for small values of $E_{\text{ms}}$; however, $\Delta M$ approaches a constant plateau (for $E_{\text{ms}}\sim 10^{-1}$) as the mainshock energy $E_{\text{ms}}$ increases. This plateau is statistically indistinguishable from the B{\aa}th's law predictions. To our knowledge, it is the first time that this law is studied for acoustic emissions experiments. The ETAS (epidemic type aftershock sequence) model presents a similar behavior for $\langle \Delta M \rangle$~\cite{Helmstetter_intro2} and only for $E_{\text{ms}}\sim10^4$ ($0.8<\beta<1.0$ in the model) that $\langle \Delta M \rangle\approx 1.2$. Empirical observations of the B{\aa}th's law for earthquakes usually report large fluctuations for $\Delta M$ estimated from individual aftershock sequences~\cite{Hainzl}. Furthermore, averaged values of $\Delta M$ also show deviations of the B{\aa}th's predictions for $E_{\text{ms}}\lesssim 10^{4}$~\cite{Gu}, which are associated to lower magnitude cutoffs in earthquake catalogs --- a situation that cannot be ruled out in our experiment.

We have thus presented an extensive characterization of the acoustic emissions of charcoal samples dampened with ethanol, aiming to establish a parallel between seismic laws { and a fracture experiment where the loading (caused by absorption and evaporation of ethanol) is internal, transient and irregular. We have found that the most fundamental seismic laws are, in general, valid in our experiments. However, some discrepancies with the case of earthquakes and fracture experiments under constant and external loading were also observed, nominally: a smaller Gutenberg-Richter exponent, a stationary behavior in the aftershocks/foreshocks rates for long times and a double power-law relationship in the productivity law. We believe that the main cause of these discrepancies is the different loading condition in our experiment. The internal stresses are irregularly distributed across the samples and may create several cracking sites acting approximately independently. This possibility partially explains the observed discrepancies: simultaneous events yield large values of energy, which contribute for a longer tail in the energy distribution and for a small power-law exponent; cracking sites operating independently also corroborates with the aleatory behavior observed long times in the aftershocks/foreshocks rates, the Omori's decay is also shorter for large values of the mainshock energies (Fig. S6); finally, the crossover behavior observed for large values of mainshock energies in the productive law can result from a superposition of mainshock events. Another possibility is that the hydrated ethanol may promote environmentally-assisted crack growth (stress corrosion) that also lead to acoustic emissions. We find very hard to direct verify these possibilities; however, stress corrosion usually happens in longer time scales (compared with our experiment), which would produce a small contribution for the acoustic events. We further believe that the simplicity of our experiment may trigger direct investigations related to previous  discussions  as well as the study of different solvents, sample sizes and charcoal materials.}

\bibliographystyle{plainnat}

\clearpage

\clearpage
\widetext
\begin{center}
\textbf{\large Supplemental Materials}
\end{center}

\setcounter{figure}{0}
\makeatletter 
\renewcommand{\thefigure}{S\@arabic\c@figure}
\renewcommand{\thetable}{S\@arabic\c@table}

\begin{figure*}[!ht]
\includegraphics[scale=0.35]{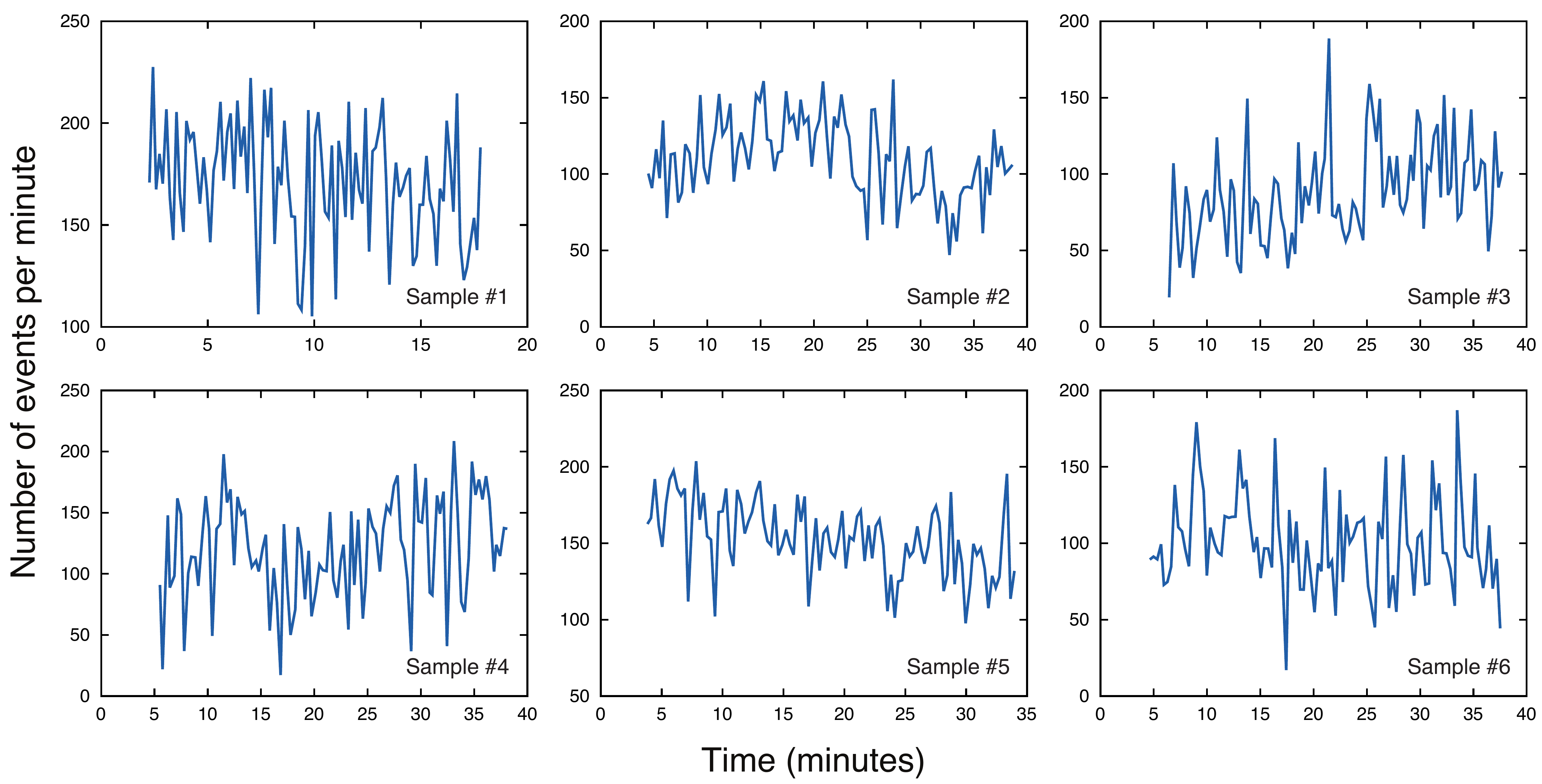}
\caption{Number of events per minute (activity rate) for each sample. Notice that all activity rates are nearly stationary, exhibiting no clear increasing or decreasing tendency.
}
\label{sfig:1}
\end{figure*}

\begin{figure*}[!ht]
\includegraphics[scale=0.35]{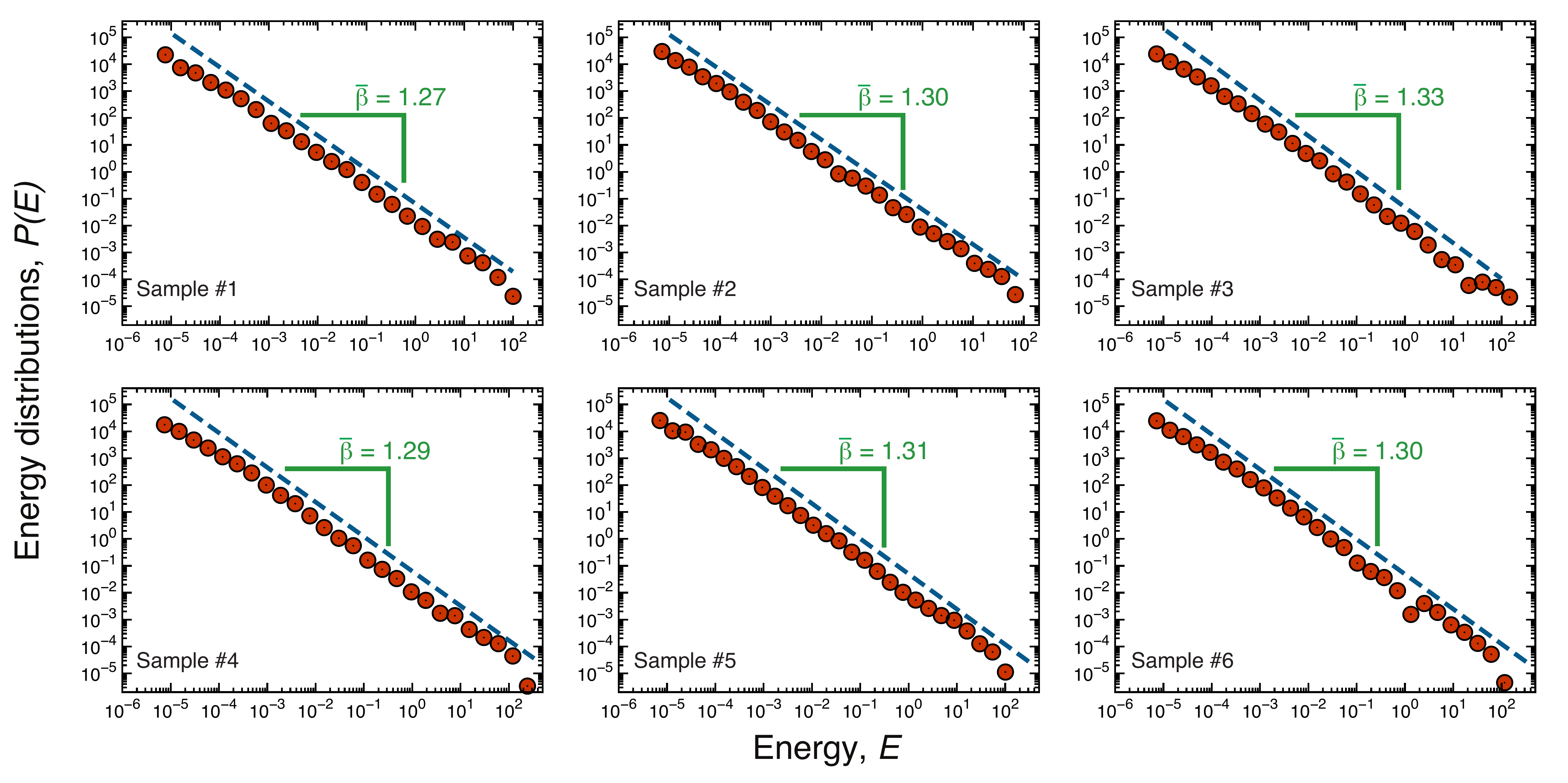}
\caption{Probability distributions of the energies $E$ for each sample. Each dashed line is a power-law decay with the power-law exponent $\bar{\beta}$ shown in the plots. Notice that all samples display approximately the same behavior. 
}
\label{sfig:2}
\end{figure*}

\begin{figure*}[!ht]
\includegraphics[scale=0.35]{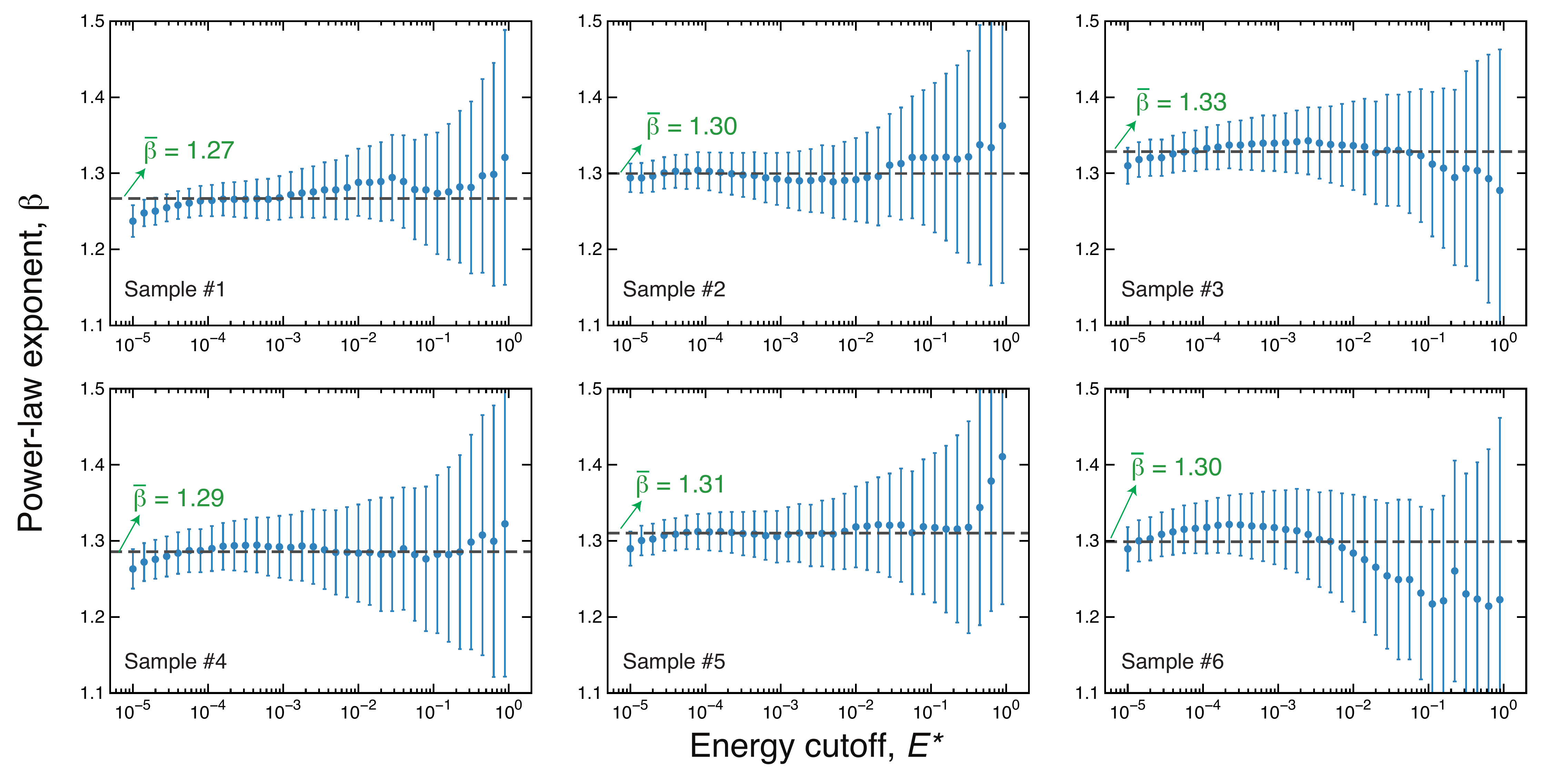}
\caption{Power-law exponents $\beta$ obtained via maximum likelihood method as a function of a lower energy cutoff $E^*$ for each sample. The error bars are $95\%$ bootstrap confidence intervals. The values of $\bar{\beta}$ are the weighted average (with weights inversely proportional to the lengths of the confidence intervals) of $\beta$ for each sample. Notice that the exponents $\beta$ are quite stable over $E^*$ for all samples.
}
\label{sfig:3}
\end{figure*}

\begin{figure*}[!ht]
\includegraphics[scale=0.35]{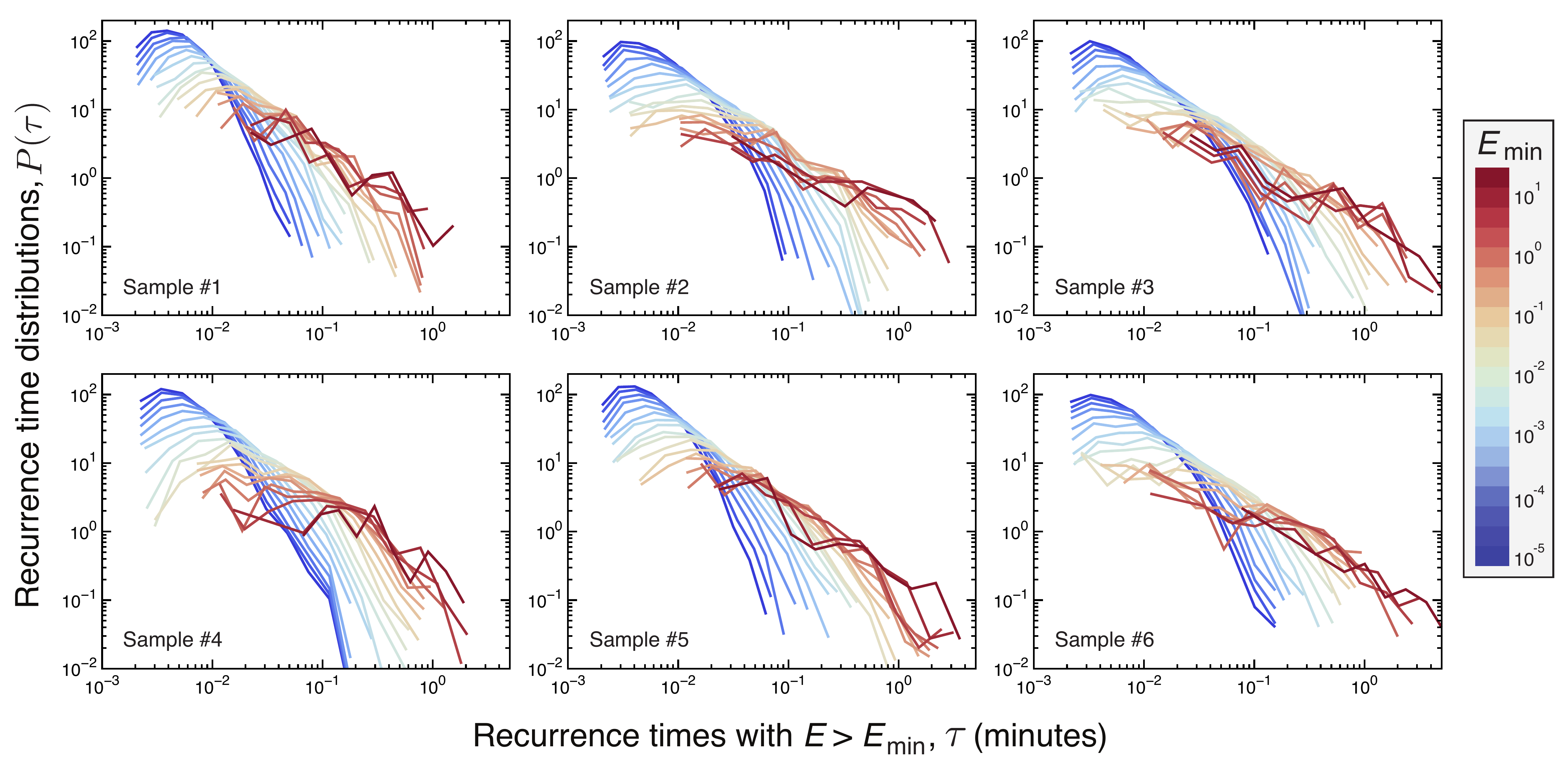}
\caption{Probability distributions of the recurrence times $\tau$ with $E>E_{\text{min}}$ for each sample. Each curve in each panel is associated with a value of $E_{\text{min}}$, as indicated by the color code. Notice that all samples display approximately the same behavior. 
}
\label{sfig:4}
\end{figure*}

\begin{figure*}[!ht]
\includegraphics[scale=0.35]{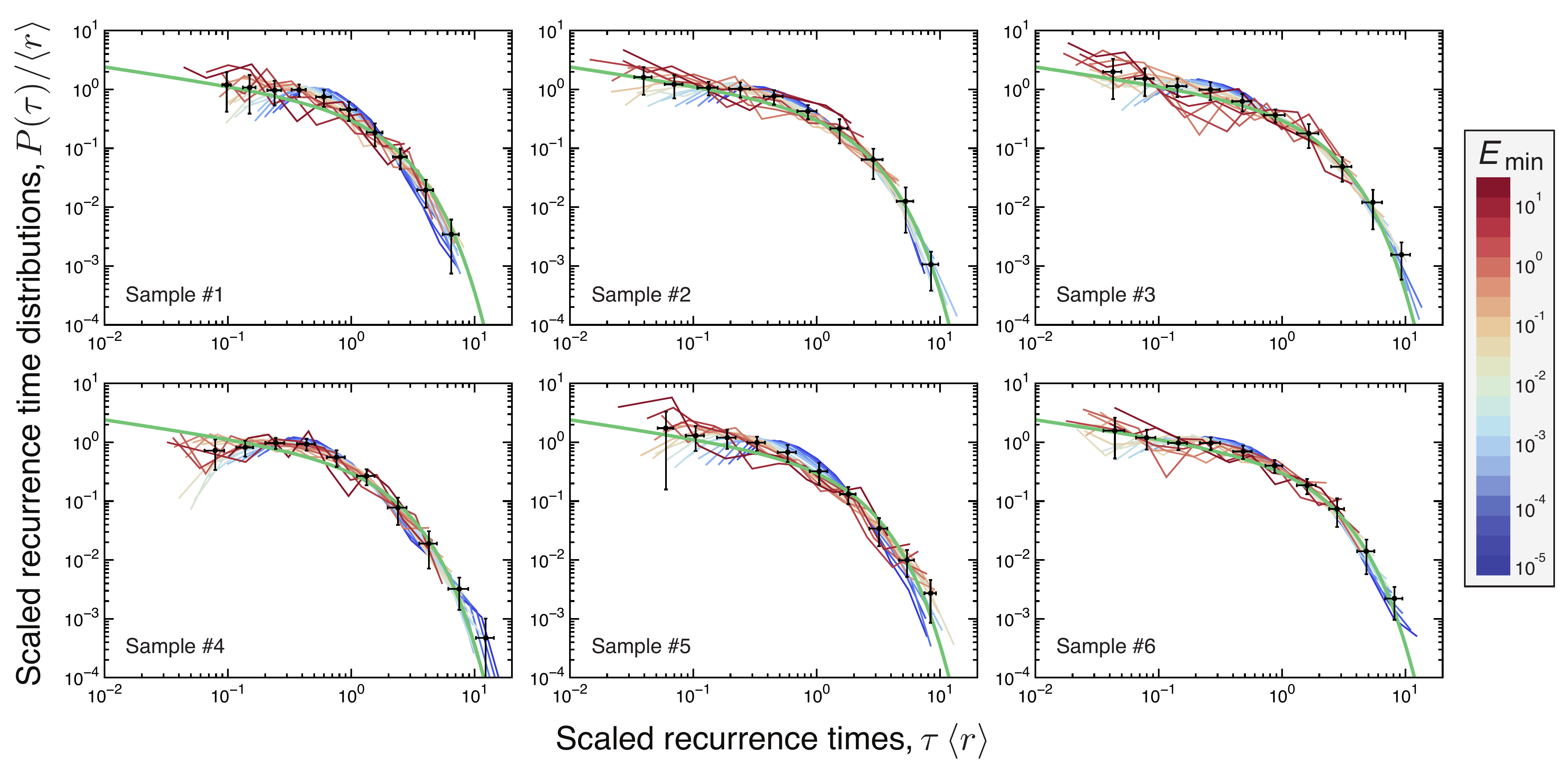}
\caption{Probability distributions of the recurrence times rescaled by the mean rate of activity $\langle r\rangle$ for different values of $E_{\text{min}}$ (indicated by the color code) for each sample. In all cases, we observe a good collapse of the distributions. The black circles are the average over all distributions for each sample and the error bars are $95\%$ bootstrap confidence intervals. The solid green lines are the gamma distribution with same parameters employed in Fig.~3b ($\gamma=0.69\pm0.08$ and $b=1.50\pm0.12$).
}
\label{sfig:5}
\end{figure*}

\begin{figure*}[!ht]
\includegraphics[scale=0.35]{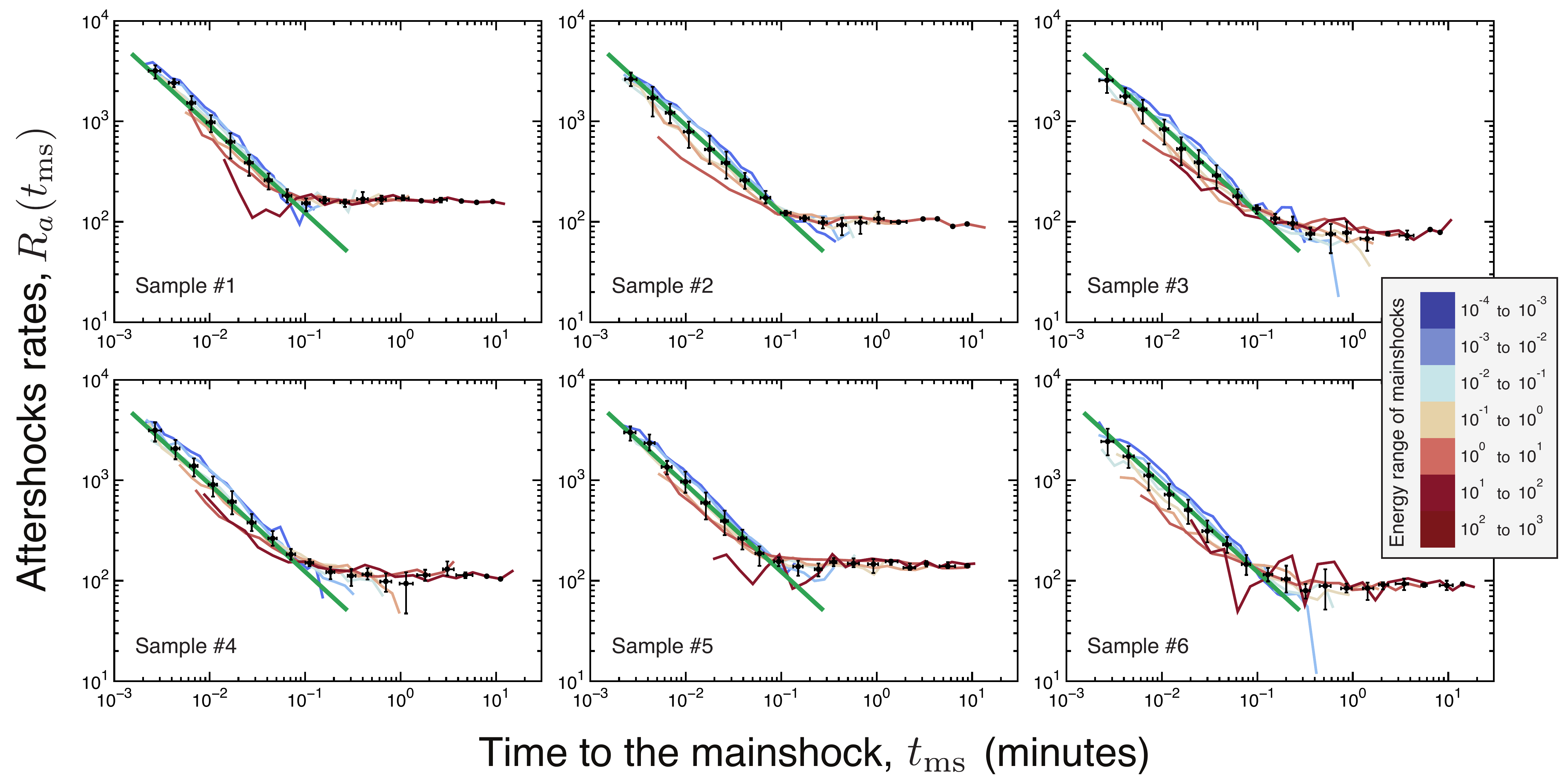}
\caption{Number of aftershocks per unit of time, $R_a(t_{\text{ms}})$, as function of time to the mainshock, $t_{\text{ms}}$, for each sample. The mainshocks have been defined as events with energy in the range $10^{j}$--$10^{j+1}$, with $j=-4,-3,\dots,3$, and color code indicates the rage employed. In all plots, the black circles are the average over all curves of the panel and the error bars are $95\%$ bootstrap confidence intervals. The black lines show the same power-law function employed in Fig.~4a, that is, $R_a(t_{\text{ms}})\sim t_{\text{ms}}^{-p}$, with $p=0.87\pm0.01$. We note the power-law regime is small for large values of mainshocks energies.
}
\label{sfig:6}
\end{figure*}

\begin{figure*}[!ht]
\includegraphics[scale=0.35]{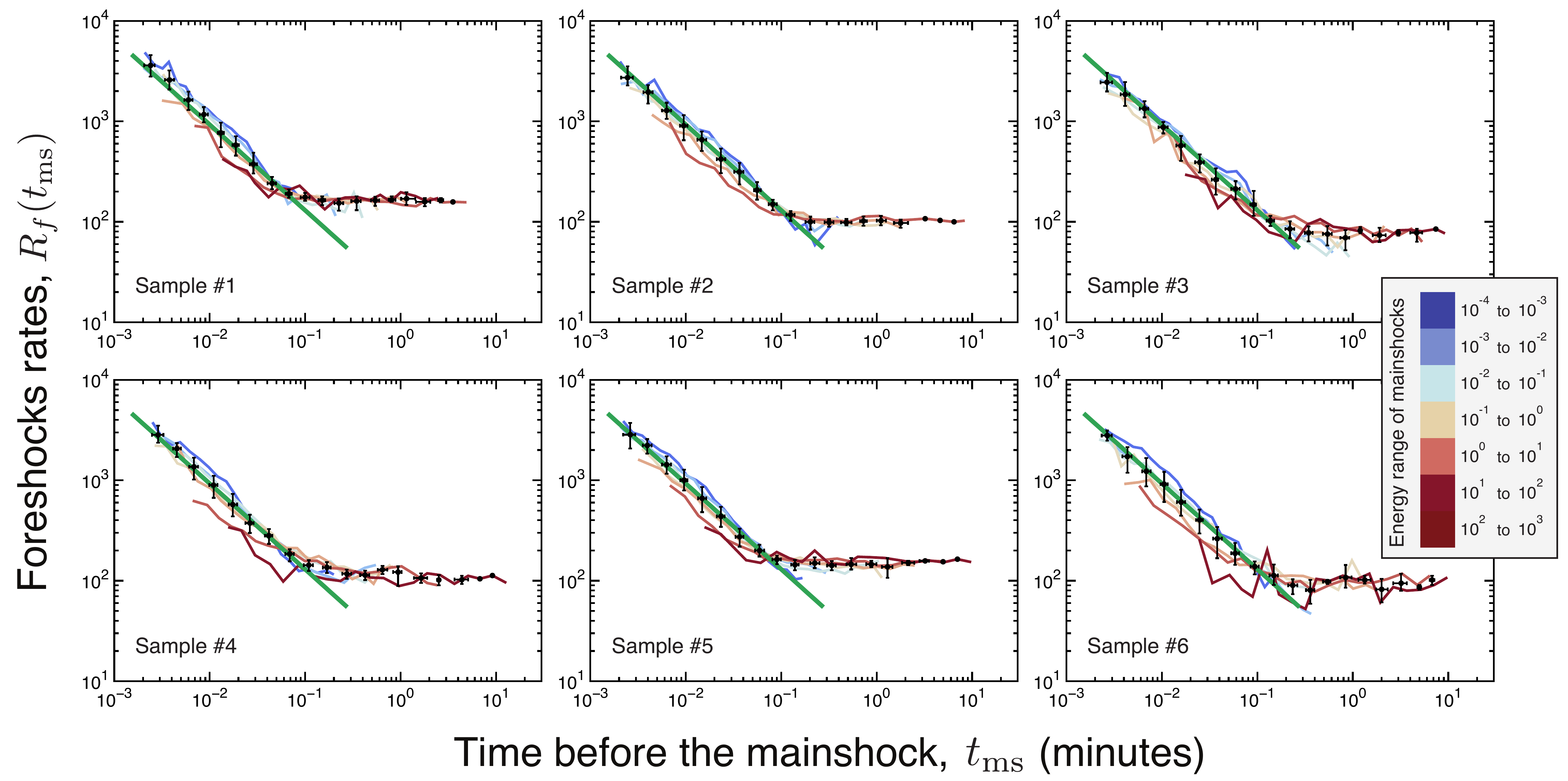}
\caption{Number of foreshocks per unit of time, $R_f(t_{\text{ms}})$, as function of time before the mainshock, $t_{\text{ms}}$, for each sample. The mainshocks have been defined as events with energy in the range $10^{j}$--$10^{j+1}$, with $j=-4,-3,\dots,3$, and color code indicates the rage employed. In all plots, the black circles are the average over all curves of the panel and the error bars are $95\%$ bootstrap confidence intervals. The black lines show the same power-law function employed in Fig.~4b, that is, $R_f(t_{\text{ms}})\sim t_{\text{ms}}^{-p}$, with $p=0.85\pm0.01$.
}
\label{sfig:7}
\end{figure*}

\begin{figure*}[!ht]
\includegraphics[scale=0.35]{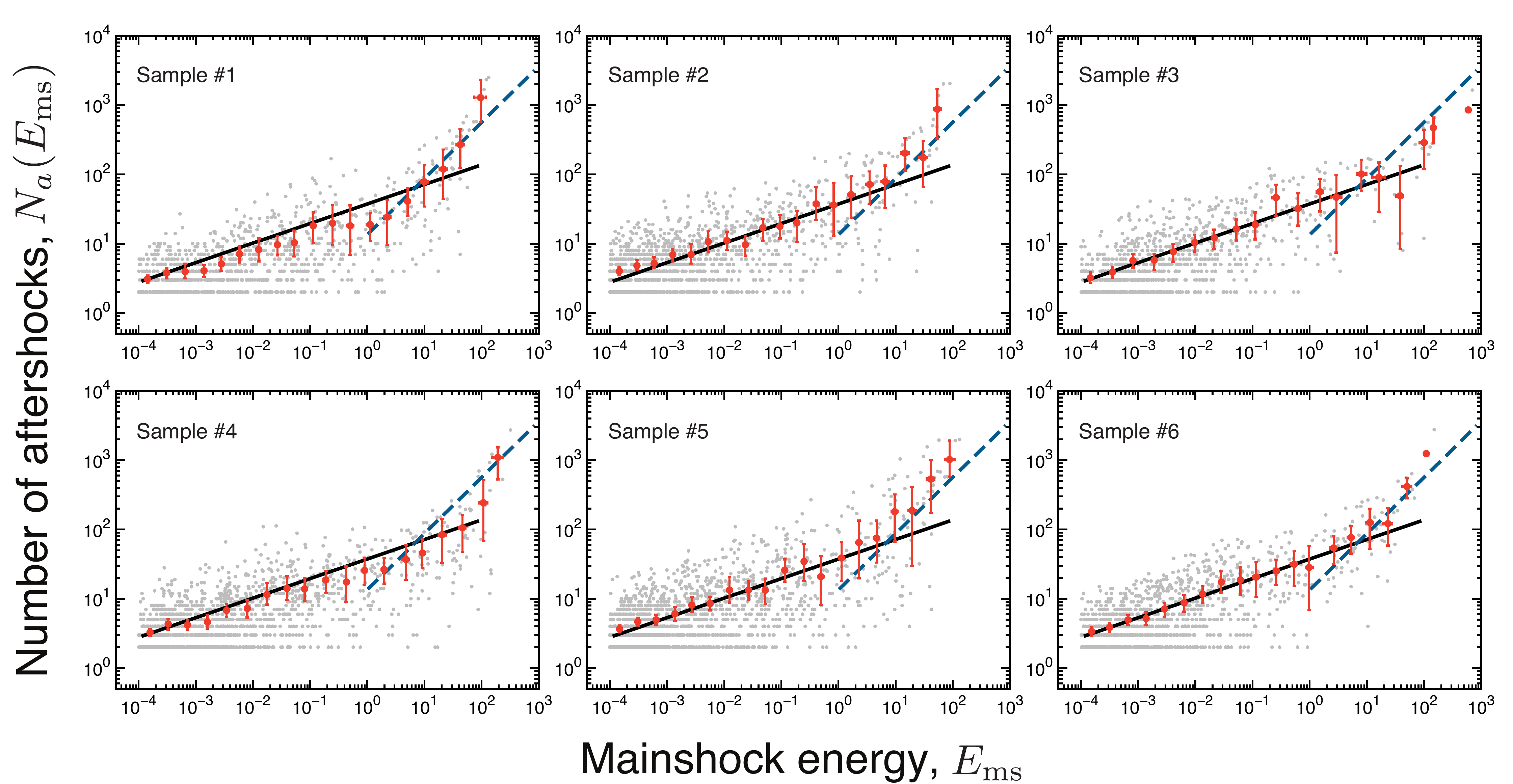}
\caption{Number of aftershocks $N_{a}(E_{\text{ms}})$ triggered by a mainshock of energy $E_{\text{ms}}$ in each sample (gray dots). In each panel, the red circles are window average values and the error bars are $95\%$ bootstrap confidence intervals. The black solid lines are the same power-law relationship, $N_{a}(E_{\text{ms}})\sim E_{\text{ms}}^\alpha$, with $\alpha=0.28\pm0.01$, shown in Fig.~5a as well as the blue dashed lines represent a power-law with $\alpha=0.81\pm0.06$.
}
\label{sfig:8}
\end{figure*}

\begin{figure*}[!ht]
\includegraphics[scale=0.35]{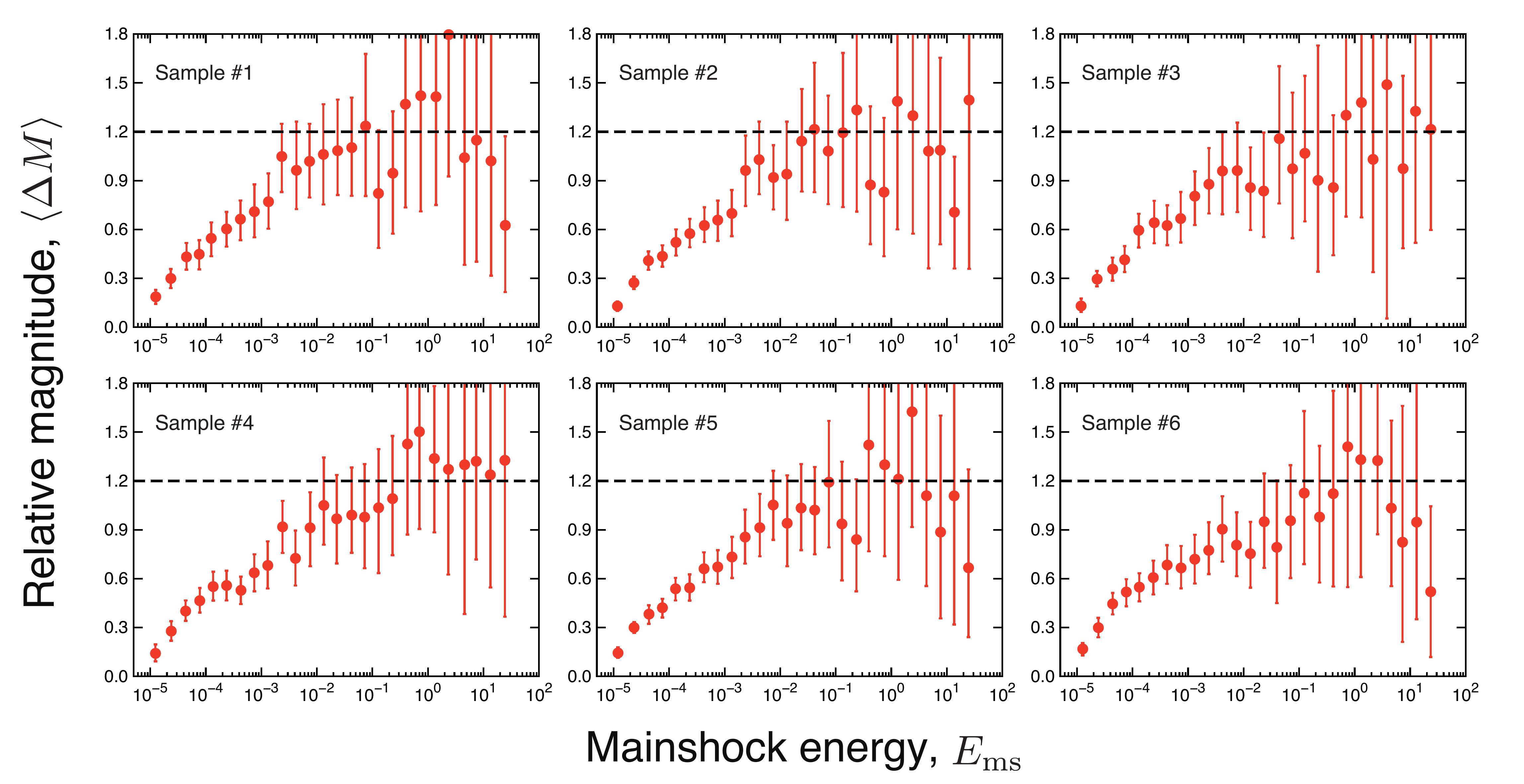}
\caption{The red circles show the average values of the relative difference in magnitude ($\langle \Delta M\rangle$) between the mainshock, $\log E_{\text{ms}}$, and its largest aftershock, $\log E_{\text{la}}$, as a function of mainshock energy $E_{\text{ms}}$ (for each sample). The horizontal axis is in log scale. The error bars are 95\% bootstrap confidence intervals. Notice that, for each sample, these relative differences approach the value of $1.2$ (dashed lines) as $E_{\text{ms}}$ increases.
}
\label{sfig:9}
\end{figure*}

\end{document}